\documentclass[11pt]{article}
\usepackage[margin=2.5cm]{geometry}
\usepackage[utf8]{inputenc}
\usepackage{amsmath,amsthm}              
\usepackage{hyperref}
\usepackage{graphicx}
\usepackage{colortbl}
\usepackage{multirow}
\usepackage{array}
\usepackage{authblk}
\usepackage{todonotes}
\usepackage{lineno}
\usepackage{float}
\usepackage{graphicx}
\usepackage{subfig}
\usepackage{tcolorbox}
\usepackage{bm}
\usepackage{mathtools}
\usepackage{commath}
\usepackage[detect-none]{siunitx}
\usepackage[sort&compress,numbers]{natbib}

\usepackage{array}
\newcolumntype{L}[1]{>{\raggedright\let\newline\\\arraybackslash\hspace{0pt}}m{#1}}
\newcolumntype{C}[1]{>{\centering\let\newline\\\arraybackslash\hspace{0pt}}m{#1}}
\newcolumntype{R}[1]{>{\raggedleft\let\newline\\\arraybackslash\hspace{0pt}}m{#1}}

\newtheorem{theorem}{Theorem}[section]
\newtheorem{lemma}{Lemma}[section]

\newtheorem*{remark}{Remark}

\def\r0{\mathcal{R}_0}

\title{\textbf{An age-of-infection model with both symptomatic and asymptomatic infections}}
\author[1]{Fan Bai\thanks{bai@hcm.uni-bonn.de}}
\affil[1]{Hausdorff Center for Mathematics, University of Bonn, Bonn, Germany}

\begin{document}
\maketitle

\medskip


\medspace


\medskip


\section{Abstract}
We formulate a general age-of-infection epidemic model with two pathways: the symptomatic infections and the asymptomatic infections. We then calculate the basic reproduction number $\mathcal{R}_0$ and establish the final size relation. It is shown that the ratio of accumulated counts of symptomatic patients and asymptomatic patients is determined by the symptomatic ratio $f$ which is defined as the probability of eventually becoming symptomatic after being infected. We also formulate and study a general age-of-infection model with disease deaths and with two infection pathways. The final size relation is investigated, and the upper and lower bounds for final epidemic size are given. Several numerical simulations are performed to verify the analytical results. 

\medspace

\section{Introduction}
Asymptomatic infections are recorded for COVID-$19$ (Coronavirus SARS-CoV-$2$) pandemic and have been studied from many different aspects. Evidence suggests that asymptomaticity has significant impacts on the development of COVID-$19$ (\cite{Rothe2020,McCulloh2020}). Furthermore, because of its "silently transmissible" nature, the pathway of asymptomatic infection is considered as a key component for understanding the mechanisms of transmission and evaluating the efficiency of intervention measures. However, quantification of asymptomatic infections and further inference of value of symptomatic ratio $f$ is tricky due to the unreliable recorded numbers of symptomatic patients and asymptomatic patients at any specific time post epidemic outbreak. For COVID-$19$, it is suggested that the asymptomatic ratio is around $35.1\%$ (\cite{Sah2021}) and it is highly age dependent. Other studies based on daily incidence data from certain geographic locations and using various parameter estimation techniques indicate very different values of symptomatic/asymptomatic ratio (e.g., \cite{McCulloh2020,Nishiura2020,Lietal2021}). The purpose of this paper is to formulate the epidemic models with two infection pathways, by considering different formats of the symptomatic ratio $f$ (either a constant or a piece-wise function), and obtain the relation between the ratio, the basic reproduction number and the final epidemic size. Another focus is to compare the final sizes of symptomatic infections and asymptomatic infections.

The Age-of-infection model was originally derived by Kermack and McKendrick in their well-known paper \cite{Kermack1927}, it is the general form of epidemic models (\cite{Brauer2005,Brauer2005age,Brauer2012,Brauerage2008,CYFB2008,Feng2007}). This type of epidemic model is especially useful to study the epidemic models in which periods are far from exponential distributed, e.g., the current COVID-$19$ pandemic. The formulation of the basic reproduction number for the age-of-infection model was well studied in \cite{CYFB2008} and the final size relation has been established in \cite{Brauerage2008}. The age-of-infection model with disease-induced mortality was also studied in \cite{Brauerage2008}, an inequality of final size relation was developed. The general model was also extended for the heterogeneous mixing population in \cite{Brauer2009}. In this paper, we formulate and study an age-of-infection model for epidemics with both symptomatic infections and asymptomatic infections. The derivation and comparison of the basic reproduction number $\mathcal{R}_0$ and the formulation of final size relations are the main focus of the study. We also consider the age-of-infection model with two infection pathways and with disease-induced mortality.

This paper consists of five parts. In section \ref{sect2}, we firstly review the basic general age-of-infection model in a homogeneously mixing population. In section \ref{sect3}, we formulate the age-of-infection model with two pathways, the symptomatic infection and the asymptomatic infection.  We then calculate the basic reproduction number and obtain the corresponding final size relation. We will focus on the ratio of total symptomatic cases and total asymptomatic cases in this study, which entirely depends on the symptomatic ratio $f$. As an example, we study a special case of Susceptible-Exposed-Infectious-Asymptomatic-Removed (SEIAR) model, by assuming exponentially distributed disease stages. In section \ref{sect4}, we further consider the disease-induced deaths for the age-of-infection model with two infection pathways and investigate the final size relations. In section \ref{sect5}, we perform several sets of numerical simulations to verify our theoretical results. In section \ref{sect6}, we summarize the investigation and also propose some interesting future work in the direction of age-of-infection modeling.

\medspace

\section{The general age-of-infection model}\label{sect2}
We have $S(t)$ denote the number of susceptibles at time $t$ and have $\phi(t)$ be the total infectivity at time $t$. $\phi(t)$ is defined as the sum of products of the number of infected members with each infection age and the mean infectivity for that infection age (\cite{CYFB2008}). We assume that on average each member of the population make $a$ contacts per unit time. The population size is $N$ and there are no demographic effects (births, deaths, migration) in the population. We further define $B(\tau)$ the fraction of infected members remaining infected at infection age $\tau$ and $\pi(\tau)$ the mean infectivity at infection age $\tau$. $B(\tau)$ is a non-increasing function and satisfies that
\begin{equation*}
B(0) = 1, \quad B(\infty) = 0.
\end{equation*}
The general age-of-infection model is 
\begin{equation}\label{ageofinfectionmodel}
	\begin{aligned}
		S' &= -\frac{a}{N}S\phi \\
		\phi(t) &= \phi_0(t) + 	\int_{0}^{t}\frac{a}{N}S(t-\tau)\phi(t-\tau)B(\tau)\pi(\tau)d\tau, 
	\end{aligned}
\end{equation}
where $\phi_0(t)$ is the total infectivity of members of the population who were infected at the initial time, at time $t$. The following Figure \ref{fig:flowchart1} demonstrates the dynamics of the general model. It is worth mentioning that the number of infectives is described as
\begin{equation*}
I(t) = I_0(t) + 	\int_{0}^{t}\frac{a}{N}S(t-\tau)\phi(t-\tau)B(\tau)d\tau. 
\end{equation*}
It is straightforward to derive the equation about the change of number of infectives (\cite{CYFB2008})
\begin{equation*}
I^{\prime}(t) = \frac{a}{N}S\phi + \int_{0}^{t} \frac{a}{N}S(\tau)\phi(\tau)B^{\prime}(t-\tau)d\tau,
\end{equation*}
where $B^{\prime}$ reflects the recovery of infected individuals.

\medspace

\begin{figure}[H]
	\centering
	\includegraphics[width = 0.79\textwidth]{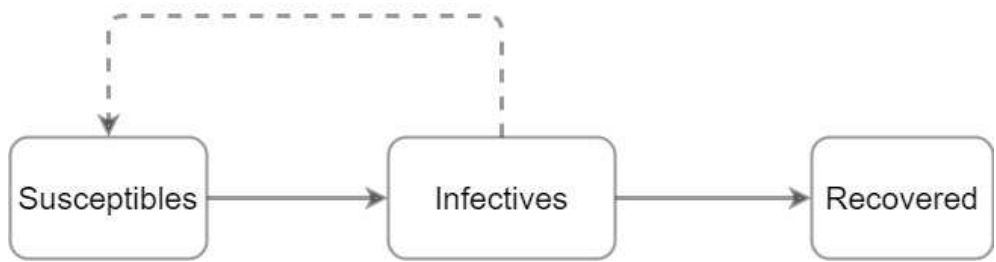}
	\caption{The diagrammatic framework of the basic age-of-infection model \eqref{ageofinfectionmodel}.}
	\label{fig:flowchart1}
\end{figure} 

\medspace

It has been well studied in \cite{Brauerage2008,CYFB2008}, that the basic reproduction number for model \eqref{ageofinfectionmodel} is
\begin{equation}\label{R0general}
	\mathcal{R}_0 = a\int_{0}^{\infty}B(\tau)\pi(\tau)d\tau.	
\end{equation}
The final size relation for model \eqref{ageofinfectionmodel} is
\begin{equation}\label{finalsize1}
	\log\frac{S_0}{S_\infty} = \mathcal{R}_0\left[1-\frac{S_\infty}{N_0}\right]-\frac{a}{N_0}\int_{0}^{\infty}[(N_0-S_0)B(t)\pi(t)-\phi_0(t)]dt.
\end{equation}
The uniqueness of $S_\infty$ has been confirmed in \cite{Brauerage2008}. If the initial infectivity is small and neglectible, the final size relation can be simplified as
\begin{equation}\label{finalsize2}
	\log\frac{S_0}{S_\infty} = \mathcal{R}_0\left[1-\frac{S_\infty}{N_0}\right].
\end{equation}

\medspace

\section{The age-of-infection model with both symptomatic and asymptomatic infections}\label{sect3}
\subsection{The model formulation}
\begin{figure}[H]
	\centering
	\includegraphics[width = 0.8\textwidth]{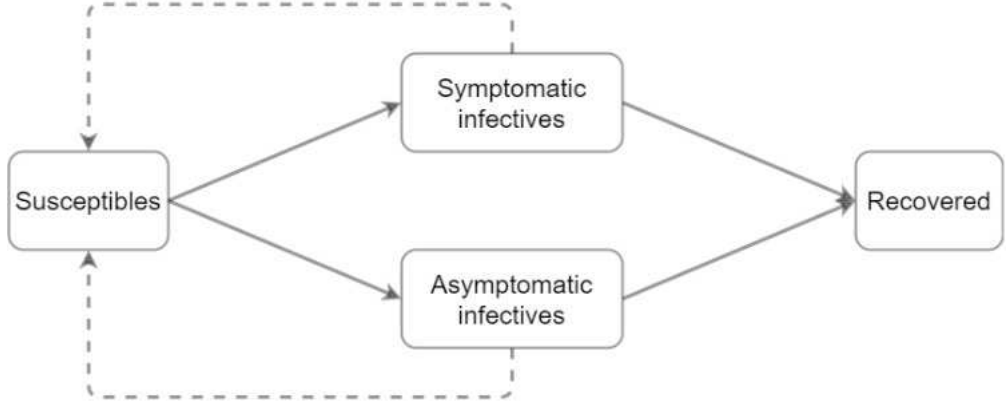}
	\caption{The diagrammatic framework of the age-of-infection model \eqref{ageofinfection}.}
	\label{fig:flowchart2}
\end{figure}

We extend the general age-of-infection model \eqref{ageofinfectionmodel} by considering that the infected individuals are possibly asymptomatic or symptomatic. It is assumed that once an individual is infected, there is a probability $0 < f < 1$ that this individual eventually becomes symptomatic, and the probability of becoming asymptomatic is $(1-f)$. We denote $\phi_i(t)$ the sum of products of the symptomatic individuals with each infection age and the mean infectivity for that infection age, $\phi_a(t)$ the sum of products of the asymptomatic individuals with each infection age and the mean infectivity for that infection age. Thus, the total infectivity is
\begin{equation}\label{relation1}
	\phi(t) = \phi_i(t)+\phi_a(t).
\end{equation}  
$B_i(\tau)$ and $B_a(\tau)$ represent the fraction of symptomatic and asymptomatic individuals remaining infected at infection age $\tau$, respectively; $\pi_i(\tau)$ and $\pi_a(\tau)$ represent the mean infectivity of symptomatic and asymptomatic individuals at infection age $\tau$, respectively. $A_i(\tau):=B_i(\tau)\pi_i(\tau)$ is the mean infectivity of a symptomatic individual at infection age $\tau$, while $A_a(\tau):=B_a(\tau)\pi_a(\tau)$ is the mean infectivity of an asymptomatic individual at infection age $\tau$. The dynamics is depicted in Figure \ref{fig:flowchart2}. It is noticed that the recovered symptomatic/asymptomatic patients are theoretically counted as one stage of the infection process, but with no infectivities.  

\medspace

\medspace
The age-of-infection model is
\begin{equation}\label{ageofinfection}
	\begin{aligned}
		S' &= - a \frac{S}{N}(\phi_i+\phi_a) \\
		\phi_i(t) &= \phi_{i,0}(t)+\int_{0}^{t}f\frac{a}{N}S(t-\tau)(\phi_i(t-\tau)+\phi_a(t-\tau))B_i(\tau)\pi_i(\tau)d\tau, \\
		\phi_a(t) &= \phi_{a,0}(t)+\int_{0}^{t}(1-f)\frac{a}{N}S(t-\tau)(\phi_i(t-\tau)+\phi_a(t-\tau))B_a(\tau)\pi_a(\tau)d\tau.
	\end{aligned}
\end{equation}
It can be proved that model \eqref{ageofinfection} is an extension of the general age-of-infection model \eqref{ageofinfectionmodel}. We first calculate the basic reproduction number for model \eqref{ageofinfection},
\begin{equation}\label{R0two}
	\mathcal{R}_0 = a \left[f\int_{0}^{\infty}A_i(\tau)d\tau+(1-f)\int_{0}^{\infty}A_a(\tau)d\tau\right].
\end{equation}
This form is consistent with Theorem $3.1$ in \cite{CYFB2008} that the basic reproduction number $\mathcal{R}_0$ depends only on the mean period in each infective stage, regardless of its distribution. We now generate the final size relation for model \eqref{ageofinfection}. We first write
\begin{equation*}
		-\frac{S'}{S} = \frac{a}{N}\left(\phi_{i,0}(t)+\phi_{a,0}(t)\right)  + \frac{a}{N}\int_{0}^{t}[-S'(t-\tau)]\left(fA_i(\tau)+(1-f)A_a(\tau)\right)d\tau.
\end{equation*}
Integration with respect to $t$ from $0$ to $\infty$ yields
\begin{equation*}
	\begin{aligned}
		\log\frac{S_0}{S_\infty} =& \frac{a}{N}\int_{0}^{\infty}\left(\phi_{i,0}(t)+\phi_{a,0}(t)\right)dt  + \frac{a}{N}\int_{0}^{\infty}\int_{0}^{t}[-S'(t-\tau)]\left(fA_i(\tau)+(1-f)A_a(\tau)\right)d\tau dt \\
		=& \frac{a}{N}\int_{0}^{\infty}\left(\phi_{i,0}(t)+\phi_{a,0}(t)\right)dt + \frac{a}{N}\int_{0}^{\infty}\left(fA_i(\tau)+(1-f)A_a(\tau)\right)\int_{\tau}^{\infty}[-S'(t-\tau)]dtd\tau \\
		=& \frac{a}{N}\int_{0}^{\infty}\left(\phi_{i,0}(t)+\phi_{a,0}(t)\right)dt  + \frac{a}{N}[S_0-S_{\infty}]\int_{0}^{\infty}\left(fA_i(\tau)+(1-f)A_a(\tau)\right)d\tau \\
		=& \frac{a}{N}\int_{0}^{\infty}\left(\phi_{i,0}(t)+\phi_{a,0}(t)\right)dt  + \mathcal{R}_0\frac{S_0-S_{\infty}}{N}. \\
	\end{aligned}
\end{equation*}
For each infection case, the probability of proceeding to the symptomatic state is $f$ and the probability of proceeding to the asymptomatic state is $(1-f)$. Thus, it is intuitive to conclude that the total number of symptomatic patients is $f(S_0-S_\infty)$ and the total number of asymptomatic patients is $(1-f)(S_0-S_\infty)$. We now perform some calculations to verity this conjecture. Firstly, the numbers of symptomatic patients $I(t)$ and asymptomatic patients $A(t)$ are
\begin{equation}\label{numbers}
\begin{aligned}
		I(t) &= I_0(t)+\int_{0}^{t}f\frac{a}{N}S(t-\tau)\phi(t-\tau)B_i(\tau)d\tau, \\
		A(t) &= A_0(t)+\int_{0}^{t}(1-f)\frac{a}{N}S(t-\tau)\phi(t-\tau)B_a(\tau)d\tau.
\end{aligned}
\end{equation} 
We now focus on $I(t)$ and calculate the derivative of $I(t)$,
\begin{equation}\label{calculateRi1}
I^{\prime}(t) = f\frac{a}{N}S\phi + \int_{0}^{t} f\frac{a}{N}S(\tau)\phi(\tau)B_i^{\prime}(t-\tau)d\tau.
\end{equation}
In Equation \eqref{calculateRi1}, the first term indicates the rate of new symptomatic infections and the second term represents the transition from infected stage to recovery stage for symptomatic patients. Thus, we have
\begin{equation*}
R_i^{\prime}(t) = -\int_{0}^{t} f\frac{a}{N}S(\tau)\phi(\tau)B_i^{\prime}(t-\tau)d\tau.
\end{equation*}
We then integrate both sides of the equation with respect to $t$ from $0$ to $\infty$,
\begin{equation}\label{result111}
\begin{aligned}
\int_{0}^{\infty}R_i^{\prime}(t)dt &= -\int_{0}^{\infty}\int_{0}^{t} f\frac{a}{N}S(\tau)\phi(\tau)B_i^{\prime}(t-\tau)d\tau dt \\
R_i(\infty) &= -\int_{0}^{\infty}\int_{\tau}^{\infty}f[-S'(t-\tau)]B_i^{\prime}(\tau)dtd\tau, \\
          &= -\int_{0}^{\infty}fB_i^{\prime}(\tau)\int_{\tau}^{\infty}[-S'(t-\tau)]dtd\tau \\
          &= -\int_{0}^{\infty}fB_i^{\prime}(\tau)[S_0-S_\infty]d\tau \\
          &=-f[S_0-S_\infty][B_i(\infty)-B_i(0)] \\
          &=f[S_0-S_\infty].
\end{aligned}
\end{equation}
Similarly, we have
\begin{equation*}
	R_a(\infty) = (1-f)[S_0-S_\infty].
\end{equation*}

\medspace

We state the following theorem to summarize all calculations. 

\begin{theorem}\label{finalsizerelation}
If the population size $N$ is large and the number of initial infection cases is small, the final size of an age-of-infection model with both symptomatic infections and asymptomatic infections is
\begin{equation}\label{finalsize}
\log\frac{S_0}{S_\infty} = \mathcal{R}_0\frac{S_0-S_{\infty}}{N},
\end{equation}  
with $\mathcal{R}_0$ being expressed in Equation \eqref{R0two}. Moreover, the total estimated cases of symptomatic infections are $f(S_0-S_\infty)$, while the total estimated cases of asymptomatic infections are $(1-f)(S_0-S_\infty)$.
\end{theorem}

If the initial infection cases are not neglectible, we have the following inequality,
\begin{equation}\label{finalsizeineq}
\log\frac{S_0}{S_\infty} > \mathcal{R}_0\frac{S_0-S_{\infty}}{N}.
\end{equation}

\medspace
It is possible that the ratio $f$ is not a constant, but a function about $t$. This may be caused by the mutations of responsible viruses. The model then becomes
\begin{equation}\label{ageofinfectionvar}
	\begin{aligned}
		S' &= - a \frac{S}{N}(\phi_i+\phi_a) \\
		\phi_i(t) &= \phi_{i,0}(t)+\int_{0}^{t}f(\tau)\frac{a}{N}S(t-\tau)(\phi_i(t-\tau)+\phi_a(t-\tau))A_i(\tau)d\tau, \\
		\phi_a(t) &= \phi_{a,0}(t)+\int_{0}^{t}(1-f(\tau))\frac{a}{N}S(t-\tau)(\phi_i(t-\tau)+\phi_a(t-\tau))A_a(\tau)d\tau.
	\end{aligned}
\end{equation}
The basic reproduction number $\mathcal{R}_0$ for model \eqref{ageofinfectionvar} depends on initial ratio $f(0)$ and is given by,
 \begin{equation}\label{R0twovar}
	\mathcal{R}_0 = a \left[f(0)\int_{0}^{\infty}A_i(\tau)d\tau+(1-f(0))\int_{0}^{\infty}A_a(\tau)d\tau\right].
\end{equation} 
$f(t)$ may take the form of a Heaviside function (with one or multiple switch times) or its smooth approximations (generalized logistic functions). Explicit final size relation for model \eqref{ageofinfectionvar} is impossible to obtain, but can be approximated by Equation \eqref{finalsize1} in Theorem \ref{finalsizerelation} if $\max(f(\tau))-\min(f(\tau))$ is small or the switch time(s) is late or close to the end of the epidemic. Some numerical simulations will be performed to show that the final sizes can not be accurately predicted, but be appropriately approximated.

\medspace

\subsection{The effect on final epidemic size estimation by using incorrectly inferred ratio}
In practice, the counts of symptomatic patients are relatively accurate. While the counts of asymptomatic patients basically rely on serology reports and surveys. Thus, it is generally a challenging task to correctly infer the value of  $f$ based on epidemiological data. This leads to an interesting problem of the comparison of outcomes by using the actual $f$ and the inaccurately inferred $f^{\prime}$. 

First we have a similar formula for $\mathcal{R}_0^{\prime}$ based on $f^{\prime}$:
\begin{equation}\label{R0twoprime}
	\mathcal{R}_0^{\prime} = a \left[f^{\prime}\int_{0}^{\infty}A_i(\tau)d\tau+(1-f^{\prime})\int_{0}^{\infty}A_a(\tau)d\tau\right].
\end{equation}
We denote
\begin{equation}\label{twodefinitions}
\begin{aligned}
	\mathcal{R}_i &:= a\int_{0}^{\infty}A_i(\tau)d\tau, \\
    \mathcal{R}_a &:= a\int_{0}^{\infty}A_a(\tau)d\tau
\end{aligned}
\end{equation}
to represent the average secondary infection cases caused by a symptomatic patient or an asymptomatic patient, when being introduced into a wholly susceptible population.
 The basic reproduction number is the linear combination of $\mathcal{R}_i$ and $\mathcal{R}_a$. The difference between $\mathcal{R}_0^{\prime}$ and $\mathcal{R}_0$ depends on the relation between ratios $f$, $f^{\prime}$, $\mathcal{R}_i$ and $\mathcal{R}_a$, 
\begin{equation}\label{comp1}
\mathcal{R}_0 - \mathcal{R}_0^{\prime} =  (f-f^{\prime})(\mathcal{R}_i - \mathcal{R}_a).
\end{equation}
Therefore, the relation between $f$ and $f^{\prime}$ is insufficient to determine how incorrect fraction will affect the assumed $\mathcal{R}_0$. It is critical to compare the values of $\mathcal{R}_i$ and $\mathcal{R}_a$. In practice, it is reasonable to assume that the inferred ratio $f^{\prime}$ is larger than the real ratio $f$, due to the fact that the asymptomatic cases are more easier to be missed and the asymptomatic infections are underestimated. The other situation is that the asymptomatic infections are overestimated, if the presymptomatic cases are incorrectly categorized as asymptomatic infections. This case indicates a smaller inferred $f^{\prime}$. However, the comparison of $\mathcal{R}_i$ and $\mathcal{R}_a$ is trivial and difficult to obtain in real life. We next introduce the established result (see Lemma $2.1$ and Lemma $2.2$ in \cite{Bai2020}) to demonstrate how different values of basic reproduction numbers will affect the final epidemic size,

\begin{theorem}\label{monoto}
The limiting susceptible population size $S_\infty$ is uniquely determined by the value of the basic reproduction number $\mathcal{R}_0$. If $\mathcal{R}_0 > 1$ and increases, $S_\infty$ decreases and the final epidemic size increases.  
\end{theorem}


\medspace

\subsection{A special case: Exponential distributions of disease stages}
\begin{figure}[H]
	\centering
	\includegraphics[width = 0.79\textwidth]{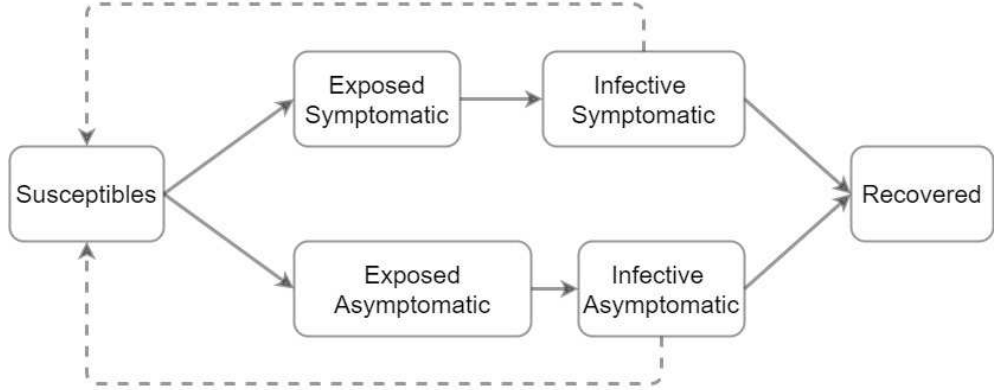}
	\caption{The diagrammatic framework of the SEIAR model \eqref{SEIAR}.}
	\label{fig:comp}
\end{figure}

A special case of the general age-of-infection model \eqref{ageofinfection} is the epidemic model with exposed stages, for both infection pathways. It is noticed that the exposed stages can be significantly different for two different types of patients. Therefore, we have the following SEIAR type of model, 
\begin{equation} \label{SEIAR}
	\begin{aligned}
		S' &= - a \frac{S}{N} (I + A)  \\
		E_i' &= af \frac{S}{N} (I + A) - \kappa_i E_i  \\
		E_a' &= a(1-f) \frac{S}{N} (I + A) - \kappa_a E_a  \\
		I' &= \kappa_i E_i - \alpha_i I  \\ 
        A' &= \kappa_a E_a - \alpha_a A  \\
		R' &= \alpha_i I + \alpha_a A.  
	\end{aligned}
\end{equation}
Now we show that model \eqref{SEIAR} can be written in the form of the general age-of-infection model \eqref{ageofinfection}. We denote $\mu_{i}(\tau)$ and $\mu_{a}(\tau)$ the fractions of symptomatic and asymptomatic patients with infection age $\tau$ who are not yet infectious, respectively, and also define $\nu_{i}(\tau)$ and $\nu_{i}(\tau)$ the fractions of symptomatic and asymptomatic patients with infection age $\tau$ who are infectious, respectively (\cite{Brauerage2008,Brauer2012}). The basic dynamics of SEIAR model \eqref{SEIAR} is described in Figure \ref{fig:comp}.

\medspace

\medspace

\begin{figure}[H]
	\centering
	\includegraphics[width = 0.8\textwidth]{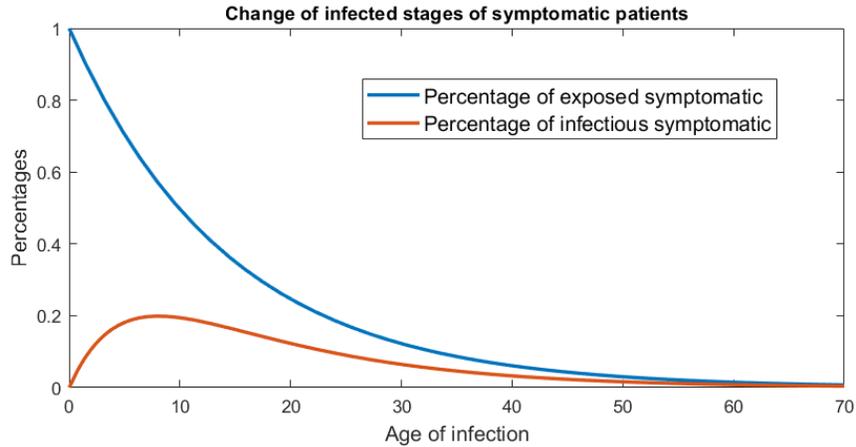}
	\caption{The change of infected stages of symptomatic patients with respect to age of infection.}
	\label{fig:stagechange}
\end{figure}

We first focus on the upper symptomatic infection path ($S\rightarrow E_i \rightarrow I \rightarrow R$) in Figure \ref{fig:comp}, $\mu_i(\tau)$ and $\nu_i(\tau)$ satisfy the ordinary differential equation
\begin{equation}\label{ODE1}
\begin{aligned}
	\mu_i^{\prime}(\tau) &= -\kappa_i \mu_i(\tau),  \\
	\nu_i^{\prime}(\tau) &= \kappa_i \mu_i(\tau) - \alpha_i \nu(\tau),
\end{aligned}
\end{equation}   
with initial condition $\mu_i(0) = 1$ and $\nu_i(0)=0$. We are able to obtain the solution of the ODE system \eqref{ODE1}
\begin{equation}\label{solution1}
	\mu_i(\tau) = e^{-\kappa_i \tau}, \quad \nu_i(\tau) = \frac{\kappa_i}{\kappa_i-\alpha_i}[e^{-\alpha_i \tau}-e^{-\kappa_i \tau}].
\end{equation}
The change of fractions $\mu_i(\tau)$ and $\nu_{i}(\tau)$ with a chosen set of parameter values ($\kappa_{i}=0.07$ and $\alpha_i=0.2$) is visualized in Figure \ref{fig:stagechange}.

The function $B_i(\tau)$ is implicitly expressed by two functions $\mu_i$ and $\nu_i$ in Equation \eqref{solution1}. It is straightforward to have the averaged infectivities for both exposed patients and symptomatic patients for $\pi_E = 0$ and $\pi_I = 1$. Based on the definition of $A_i(\tau)$, we have 
\begin{equation*}
A_i(\tau) = \frac{\kappa_i}{\kappa_i-\alpha_i}[e^{-\alpha_i \tau}-e^{-\kappa_i \tau}].
\end{equation*}
Similarly, the lower asymptomatic infection path  ($S\rightarrow E_a \rightarrow A \rightarrow R$) can be analyzed in the same manner. It implies
\begin{equation*}
	A_a(\tau) = \frac{\kappa_a}{\kappa_a-\alpha_a}[e^{-\alpha_a \tau}-e^{-\kappa_a \tau}].
\end{equation*}
From the formula for the basic reproduction number $\mathcal{R}_0$ in Equation \eqref{R0two}, we have
\begin{equation}\label{R0example}
\begin{aligned}
\mathcal{R}_0 &=  a \left[f\int_{0}^{\infty}A_i(\tau)d\tau+(1-f)\int_{0}^{\infty}A_a(\tau)d\tau\right]	 \\
&= a \left[f\int_{0}^{\infty}\frac{\kappa_i}{\kappa_i-\alpha_i}(e^{-\alpha_i \tau}-e^{-\kappa_i \tau})d\tau+(1-f)\int_{0}^{\infty}\frac{\kappa_a}{\kappa_a-\alpha_a}(e^{-\alpha_a \tau}-e^{-\kappa_a \tau})d\tau\right]	\\
&= a\left(\frac{f}{\alpha_i}+\frac{(1-f)}{\alpha_a}\right). 
\end{aligned}
\end{equation} 
This result can also be verified by using the next generation matrix approach (\cite{vandenDriessche2002}). If we consider the averaged infectivity for asymptomatic patients is $\epsilon < 1$, it can be calculated that $A_a(\tau) = 
\epsilon e^{-\kappa_i \tau} +
\frac{\kappa_a}{\kappa_a-\alpha_a}[e^{-\alpha_a \tau}-e^{-\kappa_a \tau}]$ and further we have
\begin{equation*}
	\mathcal{R}_0 = a\left(\frac{f}{\alpha_i}+\frac{(1-f)\epsilon}{\alpha_a}\right). 
\end{equation*}
The final size relation for model \eqref{SEIAR} can be generated by Theorem \eqref{finalsizerelation}. We can also use the alternative method (\cite{Brauer2008}) to verify this result. Firstly, we denote
\begin{equation*}
	f(\infty) := \lim\limits_{t \rightarrow \infty} f(t), \quad \hat{f} := \int_{0}^{\infty} f(t)dt,
\end{equation*}
if function $f$ is a non-negative integrable function defined on $0 \leq t < \infty$. It follows the lemma (see \cite{Brauer2008}),
\begin{lemma}
	If $f(t)$ is a non-negative monotone nonincreasing continuously differentiable function, then as $t \rightarrow \infty$, $f(t) \rightarrow f_{\infty} \geq 0$ and $f'(t) \rightarrow 0$. 
\end{lemma}

\medspace
Because we have
\begin{equation*}
	(S+E_i+E_a+I+A)^{\prime} = -\alpha_i I - \alpha_a A \leq 0,
\end{equation*}
this leads to $I(\infty)=0$ and $A(\infty)=0$. We further obtain the number of all infection cases
\begin{equation}\label{eq1}
	S(0)-S(\infty) = \alpha_i \hat{I} + \alpha_a \hat{A}.
\end{equation}
We then add the second equation and the fourth equation, the third equation and the fifth equation in \eqref{SEIAR}, respectively,
\begin{equation*}
\begin{aligned}
(E_i+I)^{\prime} &= af \frac{S}{N} (I + A) - \alpha_i I, \\
(E_a+A)^{\prime} &= a(1-f) \frac{S}{N} (I + A) - \alpha_i I.
\end{aligned}
\end{equation*} 
To integrate the above equations with respect to $t$ from $0$ to $\infty$ implies
\begin{equation*}
	\begin{aligned}
		E_i(\infty)+I(\infty)-E_i(0)-I(0) &= af \frac{S}{N} (\hat{I} + \hat{A})-\alpha_i\hat{I}, \\
		E_a(\infty)+A(\infty)-E_a(0)-A(0) &= a(1-f) \frac{S}{N} (\hat{I} + \hat{A})-\alpha_a\hat{A}.
	\end{aligned}
\end{equation*}
It implies the linear relation between $\hat{I}$ and $\hat{A}$,
\begin{equation}\label{eq2}
	\hat{I} = \frac{f\alpha_a}{(1-f)\alpha_i}\hat{A}.
\end{equation} 
From the first equation in model \eqref{SEIAR}, we have
\begin{equation*}
	\begin{aligned}
		-\frac{S'}{S} &= \frac{a}{N}I + \frac{a}{N}A, \\
		\log\frac{S(0)}{S(\infty)} &= \frac{a}{N}\bar{I} + \frac{a}{N}\bar{A}.
	\end{aligned}
\end{equation*}
Using equations \eqref{eq1} and \eqref{eq2}, the final size relation for model \eqref{SEIAR} can be obtained,
\begin{equation}\label{specialcase}
	\begin{aligned}
		\log\frac{S(0)}{S(\infty)} &= a\left(\frac{f}{\alpha_i}+\frac{(1-f)}{\alpha_a}\right)\frac{S(0)-S(\infty)}{N}, \\
		\log\frac{S(0)}{S(\infty)} &= \mathcal{R}_0 \frac{S(0)-S(\infty)}{N}.
	\end{aligned}
\end{equation}

It is noticed that, in Equation \eqref{eq2}, $\hat{I}$ and $\hat{A}$ can be interpreted as the sum of lengths of infection periods for all symptomatic and asymptomatic patients, respectively. If we define the length of infection period for symptomatic patient $j$ ($j \in [1,2,\cdots,\textrm{final size of symptomatic patients}] $) is $\frac{1}{\alpha_i}|_j$, the law of large numbers indicates that
\begin{equation*}
\hat{I} := \sum_{j=1}\frac{1}{\alpha_i}\bigg|_j \approx \textrm{final size of symptomatic patients} \times \frac{1}{\alpha_i}.
\end{equation*}
Same argument can be made for asymptomatic infections. Apparently the second part of the above equation holds when the final size of symptomatic patients is relatively large, and $\frac{1}{\alpha_i}$ is the pre-defined mean length of infection period. Therefore, we have the final size of symptomatic patients is $\hat{I}\alpha_i$ and the final size of asymptomatic patients is is $\hat{A}\alpha_a$. This also implies that the ratio of sizes of symptomatic and asymptomatic patients is $\frac{f}{1-f}$. In stochastic epidemic modeling with only symptomatic path (the Sellke stochastic model), a similar term $\frac{a}{N}\hat{I}$ is defined to measure the total accumulative infection pressure on a given susceptible individual during the course of the epidemic (\cite{Sellke1983,Britton2010}).

\medspace

\begin{remark}
For two pathways of symptomatic infection and asymptomatic infection, the structures (functions $B_i$ and $B_a$) can be varied significantly. The stage of pre-symptomatic in the pathway of symptomatic infection has been frequently included in modeling COVID-$19$ pandemic. For an individual who is pre-symptomatic, he or she can infect other susceptible individuals. The infectivity is likely to be even higher than the patients with symptoms (\citep{CDC2021}). Thus, for upper infection pathway in Figure \ref{fig:comp}, it is reasonable to add one more compartment "pre-symptomatic" to describe the intermediate infectious stage. The corresponding function $B_i$ can be constructed straightforwardly. It is also possible to consider that the stays in the exposed periods (or in other compartments) follow the Gamma distributions and assume multiple exposed sub-stages (\cite{Bai2018,Brauer2019book,Subramanian2021}).
\end{remark}




\medspace

\section{The age-of-infection model with disease deaths}\label{sect4}

If we consider there are disease deaths, the total population size does not remain constant. We denote $N(t)$ is the population size over time and $N_0$ is the initial population size. We further assume that a fraction $\mu_1$ ($\mu_2$) of symptomatic (asymptomatic) infectives dies of disease, and $\mu_1 > \mu_2$. The contact rate $a(N)$ is a density dependent saturating function (\cite{Brauerage2008,Brauer2012}), where $a(N)$ is a non-decreasing function about $N$ and $\frac{a(N)}{N}$ is a non-decreasing function about $N$. The age-of-infection epidemic model with disease deaths is
\begin{equation}\label{ageofinfectionwithdeath}
	\begin{aligned}
		S' &= - a(N) \frac{S}{N}(\phi_i+\phi_a) \\
		\phi_i(t) &= \phi_{i,0}(t)+\int_{0}^{t}f[-S'(t-\tau)]A_i(\tau)d\tau, \\
		\phi_a(t) &= \phi_{a,0}(t)+\int_{0}^{t}(1-f)[-S'(t-\tau)]A_a(\tau)d\tau.
	\end{aligned}
\end{equation}
Because of the disease deaths, the exact final epidemic size for model \eqref{ageofinfectionwithdeath} is impossible to obtain. As the basic reproduction number $\mathcal{R}_0$ only depends on the initial states of the epidemic and the population, it can be explicitly defined as (\cite{Brauer2008book,Brauer2005,Brauerage2008}) 
\begin{equation}\label{R0withdeaths}
	\mathcal{R}_0 = a(N_0) \left[f\int_{0}^{\infty}A_i(\tau)d\tau+(1-f)\int_{0}^{\infty}A_a(\tau)d\tau\right].
\end{equation}
We now establish the final size relation for model \eqref{ageofinfectionwithdeath} with respect to the value of $\mathcal{R}_0$ in \eqref{R0withdeaths}. Based on the properties of functions $a(N)$ and $\frac{a(N)}{N}$ (\cite{Brauer2008,Brauerage2008}), $\forall t \in [0,\infty)$, we have:
\begin{equation}\label{ineq}
	\frac{a(N_0)}{N_0} \leq \frac{a(N(t))}{N(t)} \leq \frac{a(N_\infty)}{N_\infty} \leq \frac{a(N_0)}{(1-\max\{\mu_1,\mu_2\})N_0},
\end{equation}
where $N_\infty$ is the final size of individuals who survive after the epidemic ends. We now integrate the first equation in model \eqref{ageofinfectionwithdeath} for $\frac{S^{\prime}}{S}$ and obtain
\begin{equation}\label{trans}
\begin{aligned}
\log\frac{S_0}{S_\infty} &= \int_{0}^{\infty} \frac{a(N(t))}{N(t)}(\phi_{i,0}(t)+\phi_{a,0}(t))dt 	\\	
	                     & + \int_{0}^{\infty}\frac{a(N(t))}{N(t)}\left[f\int_{0}^{t}(-S^{\prime}(t-\tau))A_i(\tau)d\tau+(1-f)\int_{0}^{t}(-S^{\prime}(t-\tau))A_a(\tau)d\tau\right]dt.
\end{aligned}
\end{equation}
Because $\frac{a(N(t))}{N(t)}$ is bounded and continuous for $t \in [0,\infty)$, we are able to find the constant $N_{\star}$ such that $N_0 \geq N_{\star} \geq N_\infty$, and satisfy
\begin{equation}\label{trans1}
	\begin{aligned}
		\log\frac{S_0}{S_\infty} =& \frac{a(N_{\star})}{N_{\star}}\bigg(\int_{0}^{\infty} (\phi_{i,0}(t)+\phi_{a,0}(t))dt 	\\	
		& + \int_{0}^{\infty}\left[f\int_{0}^{t}(-S^{\prime}(t-\tau))A_i(\tau)d\tau+(1-f)\int_{0}^{t}(-S^{\prime}(t-\tau))A_a(\tau)d\tau\right]dt\bigg)	\\
		=& \frac{a(N_{\star})}{N_{\star}}\bigg(\int_{0}^{\infty} (\phi_{i,0}(t)+\phi_{a,0}(t))dt 	\\	
		& + \int_{0}^{\infty}\left[f A_i(\tau)\int_{\tau}^{\infty}(-S^{\prime}(t-\tau))dt+(1-f)A_a(\tau)\int_{\tau}^{\infty}(-S^{\prime}(t-\tau))dt\right]d\tau\bigg) \\
		=& \frac{a(N_{\star})}{N_{\star}}\bigg(\int_{0}^{\infty} (\phi_{i,0}(t)+\phi_{a,0}(t))dt  + (S_0-S_\infty)\int_{0}^{\infty}\left(f A_i(\tau)+(1-f)A_a(\tau)\right)d\tau\bigg).	
	\end{aligned}
\end{equation} 
Similarly, if we consider both the initial symptomatic cases and initial asymptomatic cases are zero or neglectible, the integral term $\int_{0}^{\infty} (\phi_{i,0}(t)+\phi_{a,0}(t))dt$ can be omitted in equation \eqref{trans1}. The estimate of final epidemic size is
\begin{equation}\label{finalsizeestimate}
\log\frac{S_0}{S_\infty} = \frac{a(N_\star)N_0}{N_\star a(N_0)}\mathcal{R}_0 \frac{S_0-S_\infty}{N_0}.
\end{equation}
It is impossible to explicitly calculate the values of $N_\star$ and $a(N_\star)$, because $N_\star$ depends on many factors, such as the shapes of functions $A_i(t)$ and $A_a(t)$. However, we can obtain a relatively accurate estimates by employing the inequalities in \eqref{ineq},
\begin{equation}\label{finalsizeineq1}
\mathcal{R}_0 \frac{S_0-S_\infty}{N_0} \leq \log\frac{S_0}{S_\infty} \leq \frac{1}{1-\max\{\mu_1,\mu_2\}}\mathcal{R}_0\frac{S_0-S_\infty}{N_0}.
\end{equation}
It has been investigated in \cite{Bai2020} that, $S_\infty$ is uniquely determined by the value of $\mathcal{R}_0$. Indicated by Theorem \ref{monoto} (Lemma $2.1$ and Lemma $2.2$ in \cite{Bai2020}), we have the lower and upper bounds for $S_\infty$. The first part of inequality in equation \eqref{finalsizeineq1} implies that $S_\infty$ is less or equal to the final susceptible population size, when the basic reproduction number is taken as $\mathcal{R}_0$ and disease-induced deaths are not considered; the second part of inequality in equation \eqref{finalsizeineq1} shows that $S_\infty$ is larger or equal to the final susceptible population size, when the basic reproduction number is taken as $\frac{1}{1-\max\{\mu_1,\mu_2\}}\mathcal{R}_0$ and there are no disease deaths. It is clear to see that, if $\max\{\mu_1,\mu_2\} \rightarrow 0$, the elementary squeeze theorem indicates that
\begin{equation*}
\log\frac{S_0}{S_\infty} \approx \mathcal{R}_0\frac{S_0-S_\infty}{N_0}.
\end{equation*}
We still argue that, if the ratio $f$ is a constant, the number of total symptomatic cases is $f(S_0-S_\infty)$ and the number of total asymptomatic cases is $(1-f)(S_0-S_\infty)$. We have the following theorem to summarize the analysis.

\medspace

\begin{theorem}
Assume the epidemic model has both symptomatic infections and asymptomatic infections, the disease induced death rates are $\mu_1$ and $\mu_2$, respectively. If the basic reproduction number is $\mathcal{R}_0$, then the final susceptible population size $S_\infty$ has the lower bound of limiting susceptible population size with basic reproduction number $\frac{1}{1-\max\{\mu_1,\mu_2\}}\mathcal{R}_0$ and with no disease death; the upper bound is the limiting susceptible population size with basic reproduction number $\mathcal{R}_0$ and without disease deaths. The fraction of total symptomatic cases and total asymptomatic cases is approximately $\frac{f}{1-f}$.   
\end{theorem}

\medspace

This theorem is the complement of Theorem $4.1$ in \cite{Brauerage2008}, it provides the upper bound for the estimate of limiting final susceptible population size with disease deaths considered. It also indicates that when the disease induced death rates are larger, it is more challenging to accurately predict the outcome of the epidemic.

\medspace

\section{Numerical experiments}\label{sect5}
In this section, we perform two sets of numerical simulations to verify the analytical results in previous sections. The size of population is $N=10^5$ and the assumed symptomatic ratio $f=0.6$ which indicates that the probability for an infected individual becoming symptomatic is $60\%$. We further assume the contact rate $a=0.5$ (/day). Regarding the infected stages, we have $\kappa_i=0.07$ (/day) and $\alpha_i=0.2$ (/day) for symptomatic infection path; and have $\kappa_a=0.03$ (/day) and $\alpha_a=0.1$ (/day) for parallel asymptomatic infection path. It is therefore calculated that the basic reproduction number $\mathcal{R}_0 = 3.5$, and $\mathcal{R}_i=2.5$ and $\mathcal{R}_a=5$. Clearly, if the symptomatic ratio $f$ is overestimated, the estimated $\mathcal{R}_0$ is less than the actual basic reproduction number. Assumptions of standard incidence and exponentially distributed periods are made to simplify the numerical simulations.

\subsection{Epidemic models without disease deaths}
We first consider there are no disease-induced deaths for both symptomatic infections and asymptomatic infections. If the ratio $f$ is a constant, we are able to obtain the exact final epidemic size. The total numbers of symptomatic patients and asymptomatic patients during the course of the epidemic are $5796$ and $3863$, respectively. It can be calculated that the ratio of two numbers is $\frac{f}{1-f} = 1.5$. The dynamics is shown in the following Figure \ref{fig:expe1}.

	 \begin{figure}[H]
	\centering
	\includegraphics[width = 1\textwidth]{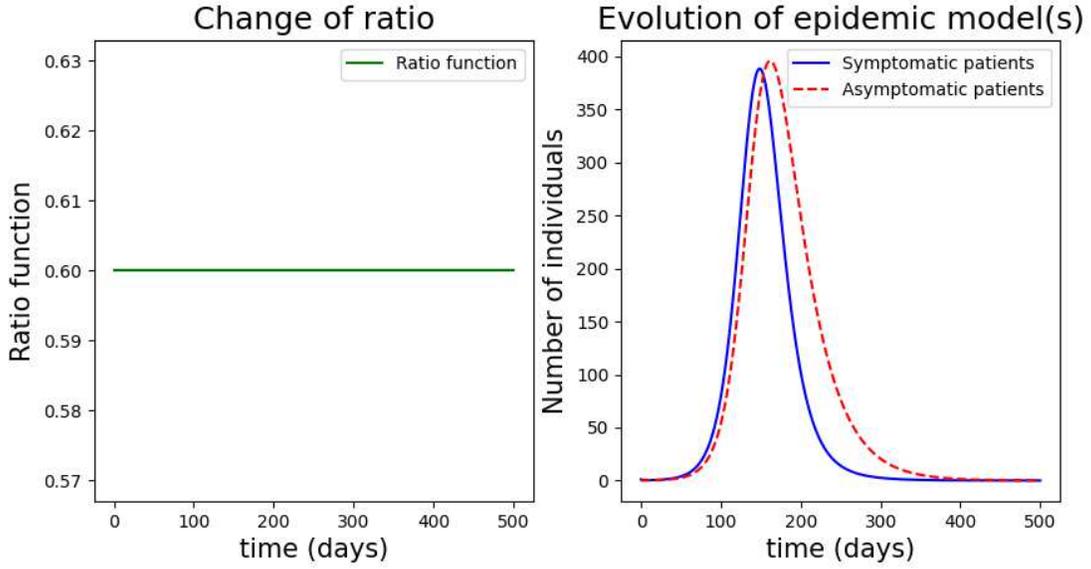}
	\caption{The constant ratio $f$ and the totally predictable progression of the epidemic.}
	\label{fig:expe1}
\end{figure}

\medspace

We next consider the case that the ratio $f$ is a function about time. Different scenarios are considered for numerical experiments:
\begin{enumerate}
	\item $f$ is a piecewise function and satisfies
	\begin{equation}\label{change1}
	f(t) =
\begin{cases}
	0.6 & \text{if $t<=150$,} \\
	0.5 & \text{if $t>150$.}
\end{cases}
	\end{equation}
	\item $f$ is a piecewise function and satisfies
	\begin{equation}\label{change2}
		f(t) =
		\begin{cases}
			0.6 & \text{if $t<=150$,} \\
			0.1 & \text{if $t>150$.}
		\end{cases}
	\end{equation}
	\item $f$ is a piece-wise function and satisfies
	\begin{equation}\label{change3}
		f(t) =
		\begin{cases}
			0.6 & \text{if $t<=150$,} \\
			0.9 & \text{if $t>150$.}
		\end{cases}
	\end{equation}
	\item $f$ is a piece-wise function and satisfies
	\begin{equation}\label{change4}
		f(t) =
		\begin{cases}
			0.6 & \text{if $t<=300$,} \\
			0.5 & \text{if $t>300$.}
		\end{cases}
\end{equation}
\end{enumerate}

\begin{figure}[H]
	\centering
	\includegraphics[width = 1\textwidth]{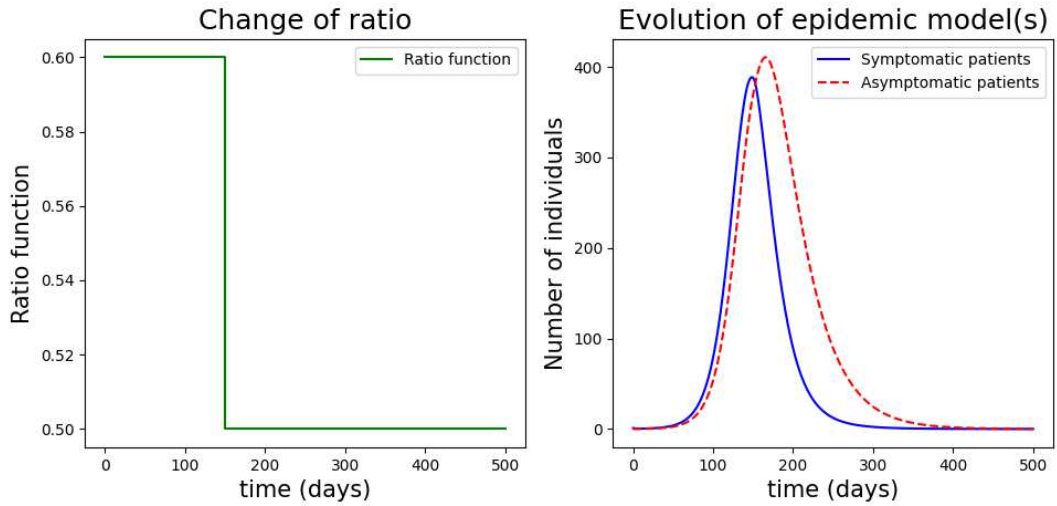}
	\caption{The time-dependent ratio $f$ in Equation \eqref{change1} and the progression of the epidemic.}
	\label{fig:expe2}
\end{figure}

\begin{figure}[H]
	\centering
	\includegraphics[width = 1\textwidth]{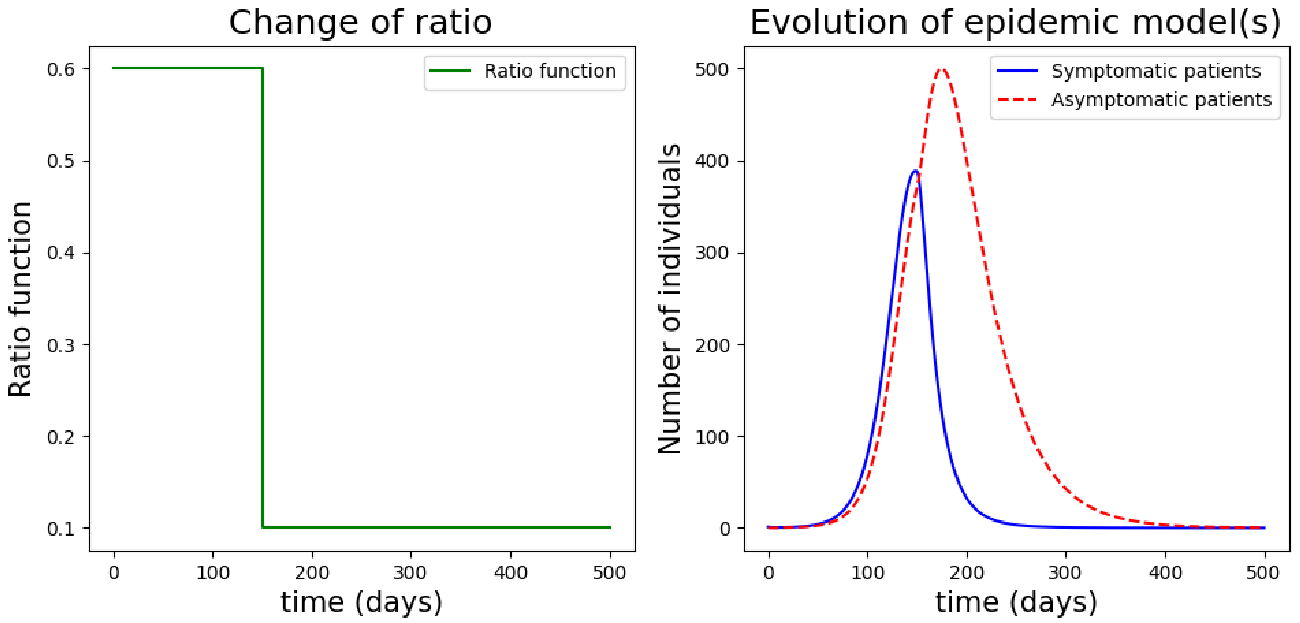}
	\caption{The time-dependent ratio $f$ in Equation \eqref{change2} and the progression of the epidemic.}
	\label{fig:expe3}
\end{figure}

\begin{figure}[H]
	\centering
	\includegraphics[width = 1\textwidth]{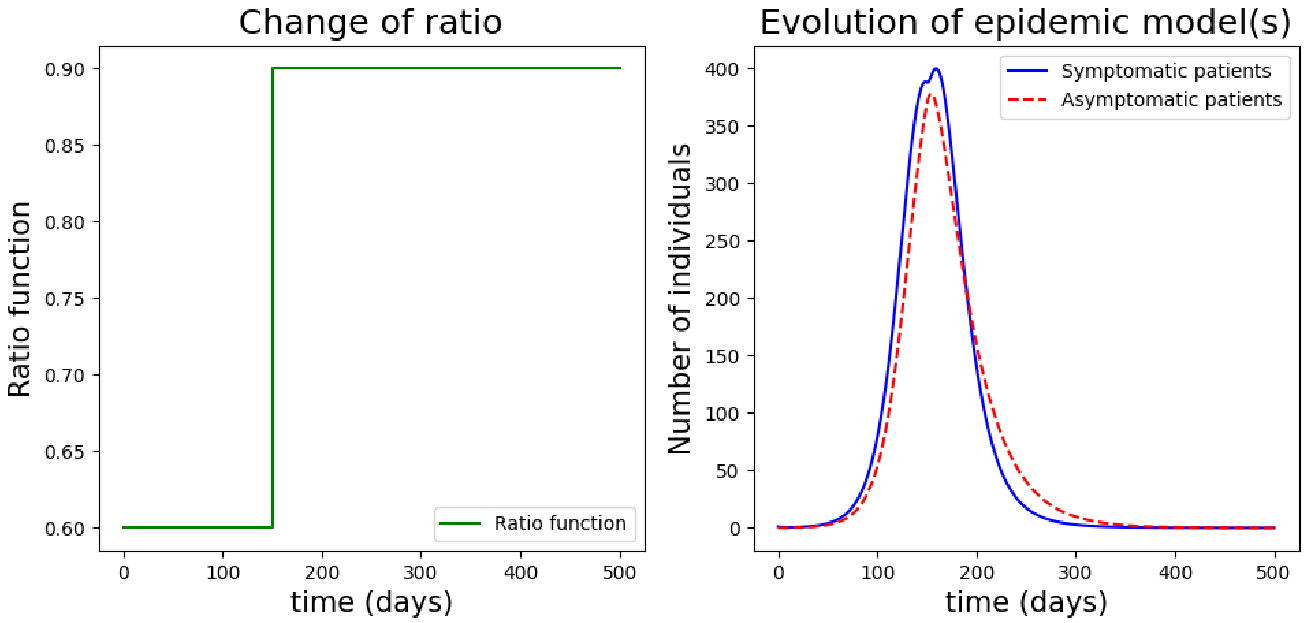}
	\caption{The time-dependent ratio $f$ in Equation \eqref{change3} and the progression of the epidemic.}
	\label{fig:expe4}
\end{figure}

\begin{figure}[H]
	\centering
	\includegraphics[width = 1\textwidth]{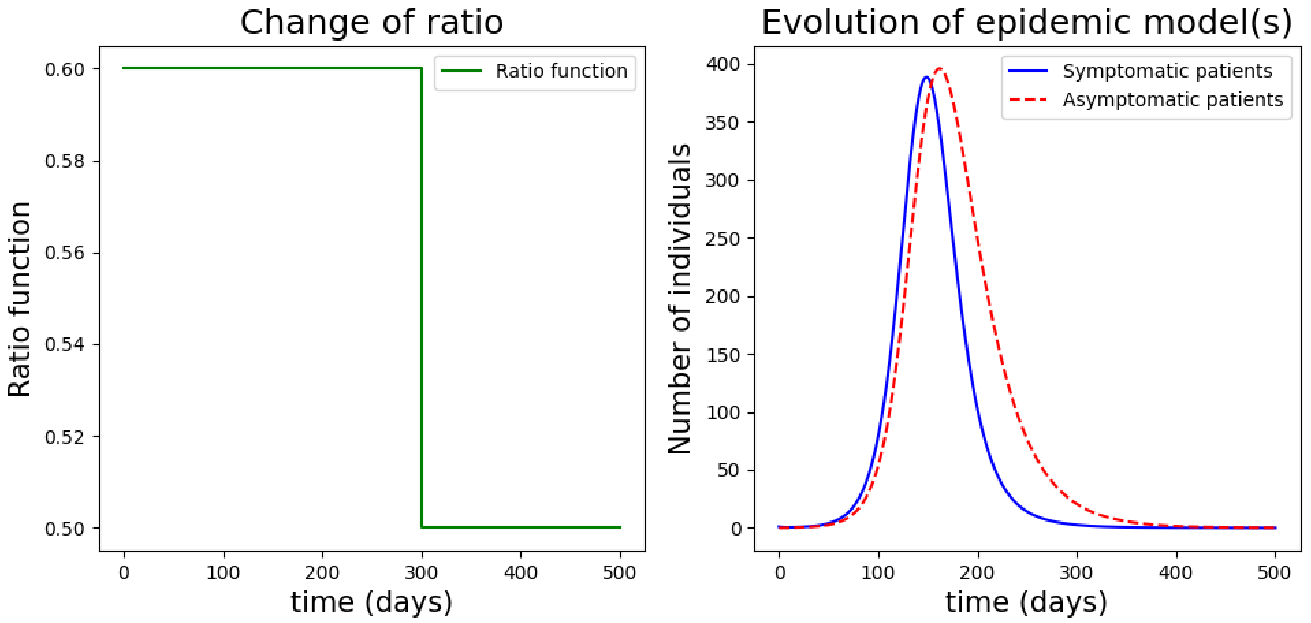}
	\caption{The time-dependent ratio $f$ in Equation \eqref{change4} and the progression of the epidemic.}
	\label{fig:expe5}
\end{figure}

The simulations are presented in Figures \ref{fig:expe2},\ref{fig:expe3},\ref{fig:expe4},\ref{fig:expe5} and the simulated outcomes are summarized in Table \ref{table1}.

\begin{table}[h!]
	\centering
		\caption{Summary of the final epidemic sizes.}
	\label{table1}
	\begin{tabular}{ |p{5cm}||p{4cm}|p{4cm}|p{2cm}|  }
		\hline
		\multicolumn{4}{|c|}{Summary of final sizes predicted for epidemic models} \\
		\hline
		Model type      & Number of total symptomatic cases  & Number of total asymptomatic cases & Ratio \\
		\hline
		Model with constant ratio       & $5796$&   $3963$  & $1.50$\\
		\hline
		Model with ratio function \eqref{change1} &  $5550$  & $4133$ & $1.34$ \\
		\hline
		Model with ratio function \eqref{change2}  & $4516$    & $5248$  & $0.86$ \\
		\hline
		Model with ratio function \eqref{change3} & $6497$ & $3081$  & $2.11$ \\
		\hline	
        Model with ratio function \eqref{change4} &  $5795$ & $3865$  & $1.49$ \\
        \hline
	\end{tabular}
\end{table}

\medskip
If we consider the simulation results for epidemic model with constant ratio as the baseline. When the switch of ratio occurs near the end of the epidemic, it is observed in Figure \ref{fig:expe5} that the final outcome can be approximated by the baseline. However, if the switch occurs earlier, the final epidemic can be significantly different and it is entirely unpredictable.

\medspace
\subsection{Epidemic models with disease deaths}
\begin{figure}[H]
	\centering
	\includegraphics[width = 1\textwidth]{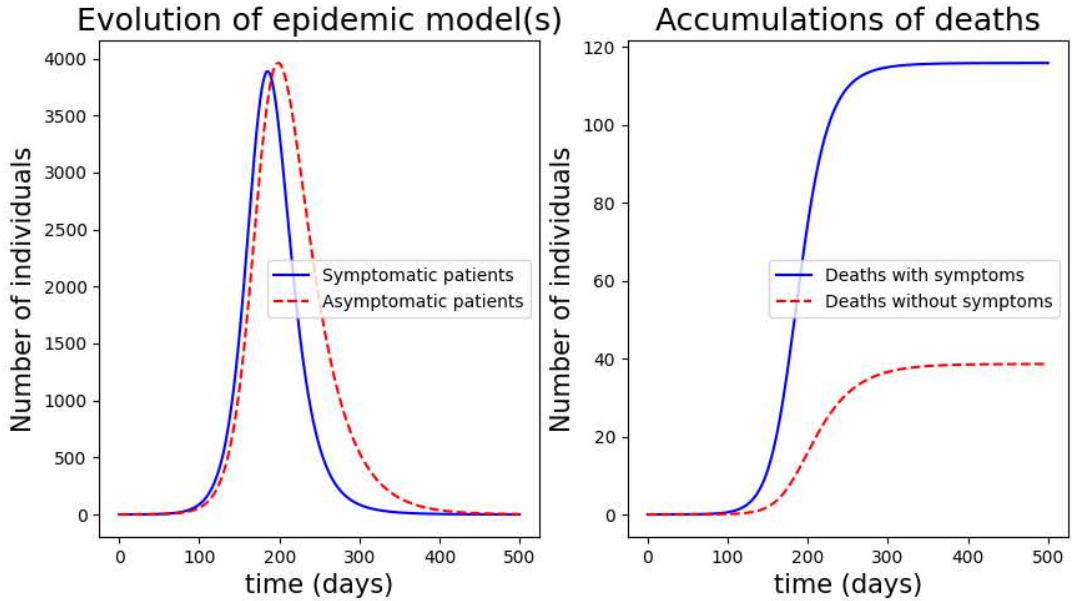}
	\caption{The progression of the epidemic with disease death rates $\mu_{1,2}=0.2\%,0.1\%$ and the accumulation of death cases.}
	\label{fig:expd1}
\end{figure}

We now consider the epidemic models with identical parameters but with a larger population size $N=10^6$ and with disease induced deaths. Firstly, we assume the death rates are $\mu_{1} = 0.2\%$ and $\mu_2 = 0.1\%$ and the dynamics of the model is simulated in Figure \ref{fig:expd1}. We are also able to obtain the number of total symptomatic cases $R_i(\infty)+D_i(\infty)=57966$ and the number of total asymptomatic cases $R_a(\infty)+D_a(\infty)=38643$. Therefore the final size of epidemic is $S(0)-S(\infty)=96609$. The ratio is $1.50$. The lower bound for final size estimation can be obtained by simulating epidemic models without disease deaths, and we have $S(0)-S(\infty)=96597$. If we consider another epidemic model without disease death and with the basic reproduction number be $\mathcal{R}_0/(1-0.2\%)$, the final epidemic size is $96623$ and this number is the upper bound for the estimation of final size for the original model with disease deaths. It can be observed that, since the death rates are very low, two bounds are both reasonable approximations.

As a comparison, we next consider the epidemic model with larger death rates $\mu_1=2\%$ and $\mu_2=1\%$. The final epidemic size is $96708$, while the upper bounds and lower bounds are $96724$ and $96597$. If the population size is larger, it becomes more difficult to accurately predict the final epidemic size.

\medspace

\section{Conclusion and future work}\label{sect6}
We have formulated an age-of-infection model to study an epidemic with symptomatic infections and asymptomatic infections. Two separate pathways are initially determined by the symptomatic ratio $f$, which is the probability that an individual being infected and eventually becoming symptomatic. The basic reproduction number $\mathcal{R}_0$ is calculated and its value depends on the symptomatic ratio $f$, the values $\mathcal{R}_i$ and $\mathcal{R}_a$. The incorrect inferred value of $f$ leads to the wrongly predicted final epidemic size. It is however a challenging task to infer these three values in practice. We then proved that, for the age-of-infection model with two infection pathways, the final epidemic size is uniquely determined by $\mathcal{R}_0$. The ratio of total numbers of symptomatic patients and asymptomatic patients is $\frac{f}{1-f}$. For the similar age-of-infection epidemic model with disease deaths, the calculation of basic reproduction number $\mathcal{R}_0$ is not affected. And the final epidemic size can not be explicitly generated, but can be approximated by the given lower bounds and upper bounds. If the death rates for both symptomatic and asymptomatic patients are smaller, the final epidemic size can be more accurately predicted. Two sets of numerical simulations were performed for two types of models.

The age-of-infection epidemic model is the general structure for different types of compartmental epidemic models (\cite{Brauer2016}) and it can be extended for other scenarios, such as the disease spread in an age-structured population with different activity levels, the impact of vaccination in the population and the behavioral changes in the population. With respect to the fundamental analysis, the complete analysis of corresponding characteristic equation for such models is still very challenging (\cite{Brauer2016,Brauer2005age,Brauer2017}).

\medspace

\noindent {\bf Declaration of Competing Interest:}
None.

\ \\
\noindent {\bf Acknowledgements:}
The work is dedicated to the author's mentor Dr. Fred Brauer ($1932$-$2021$). The author is grateful for Fred's guidance. This work is based on numerous discussions between the author and Fred. The author also acknowledges the Post-doctoral fellowship offered by Hausdorff Center for Mathematics and the University of Bonn.

\clearpage
\newpage

\clearpage
\newpage

\bibliography{age_of_infection_model_with_asymptomatic}
\bibliographystyle{plain}
\end{document}